\documentclass[aip,jap,reprint,twocolumn]{revtex4-1}
\usepackage{amsmath}
\usepackage{amssymb}
\usepackage{graphicx}
\usepackage{url}
\begin{document}
\title{Orientation-dependent magnetism and orbital structure of strained YTiO$_3$ films on LaAlO$_3$ substrates}
\author{Xin Huang}
\author{Qingyu Xu}
\author{Shuai Dong}
\email{sdong@seu.edu.cn}
\affiliation{Department of Physics, Southeast University, Nanjing 211189, China}
\date{\today}

\begin{abstract}
The strain tuned magnetism of YTiO$_3$ film grown on the LaAlO$_3$ ($110$) substrate is studied by the method of the first principles, and compared with that of the ($001$)-oriented one. The obtained magnetism is totally different, which is ferromagnetic for the film on the ($110$) substrate but A-type antiferromagnetic on the ($001$) one. This orientation-dependent magnetism is attributed to the subtle orbital ordering of YTiO$_3$ film. The $d_{xz}$/$d_{yz}$-type orbital ordering is predominant for the ($001$) one, but for the ($110$) case, the $d_{xy}$ orbital is mostly occupied plus a few contribution from the $d_{xz}$/$d_{yz}$ orbital. Moreover, the lattice mismatch is modest for the ($110$) case but more serious for the ($001$) one, which is also responsible for this contrasting magnetism.

\end{abstract}
\maketitle

%%-------------------------------------------------------------------------
%
%\section{INTRODUCTION}
Recent advances in thin-film deposition techniques have made it possible to fabricate high quality epitaxial oxide thin films and heterostructures. Consequently, it becomes a very promising route to engineer physical properties of oxide thin films by strain. One of the strain effects is that the ground state phases and phase boundaries can be tuned away from the corresponding bulk constituents, which enables us to design ``artificial" states with desired properties which are not available in bulk materials.\cite{Zubko:Arcmp,May:Nm,Dong:Prb12,Zhang:Prb12} Generally, the strain is imposed by the constraint, namely coherently grown films share the same in-plane lattice parameters with the underlying substrates.\cite{Schlom:Armr} Especially for the perovskites, due to the strain, the tilting and rotations of the oxygen octahedrons will change, which are crucial to determine the properties of perovskite oxides.\cite{Rondinelli:Am,Zhai:Nc}

For example, LaTiO$_3$ films grown on compressive substrates (e.g. LaAlO$_3$) are predicted to show the A-type antiferromagnetism (A-AFM). In contrast, those grown on tensile substrates (e.g. LaScO$_3$) maintain the G-type antiferromagnetism (G-AFM) as in bulk.\cite{Weng:Jap} It is reported that the compressive LaTiO$_3$ films may even undergo an insulator-to-metal transition.\cite{Dymkowski:Prb,He:Prb} Moreover, the strain effects depend on not only the simple lattice constants (e.g. compressive or tensile), but also the lattice orientations. It has been found that the electronic properties of films can be very different when growing along different orientations. Still taking the LaTiO$_3$ thin films as an example, the films grown on the ($001$)-oriented SrTiO$_3$ substrates show metallic behavior, while the films grown on the ($110$)-oriented DyScO$_3$ substrates are highly insulating, although the lattice mismatches are proximate for these two subtrates.\cite{Wong:Prb}

YTiO$_3$ is another interesting correlated electronic system, which is a prototypical Mott-insulator with a ferromagnetic (FM) ground state.\cite{Mochizuki:Njp} Due to the small size of Y$^{3+}$ ion, YTiO$_3$ has a highly distorted orthorhombic structure.\cite{Knofo:Prb,Komarek:Prb} From previous studies, it is known that the lattice distortions and orbital orderings are important to understand the FM order in YTiO$_3$.\cite{Knofo:Prb,Mochizuki:Njp} Theoretical studies revealed that the FM order in bulk is stabilized by the distorted Ti-O-Ti bond angles and entangled with the orbital ordering.\cite{Khaliullin:Prl} Thus, such an orbital ordering driven magnetic phase should be highly sensitive to epitaxial strain.

In this work, the epitaxial strain effects on the ground magnetic order of YTiO$_3$ film grown on the LaAlO$_3$ ($110$) substrate is studied using the first-principles method. Our calculation shows that such a YTiO$_3$ film sustains the original FM order as in the bulk constituent, which is completely different from the ($001$)-oriented case in which the film is predicted to show the A-AFM.\cite{Huang:Jap} The distortions of TiO$_6$ octahedra play an important role to determine such an orientation-dependent magnetism. In addition, the distinct lattice mismatch between YTiO$_3$ films and LaAlO$_3$ substrates for these two orientations is also responsible for this contrasting behavior.

%%------------------------------------------------------------------------------
%

\begin{figure}
\centering
\includegraphics[width=0.5\textwidth]{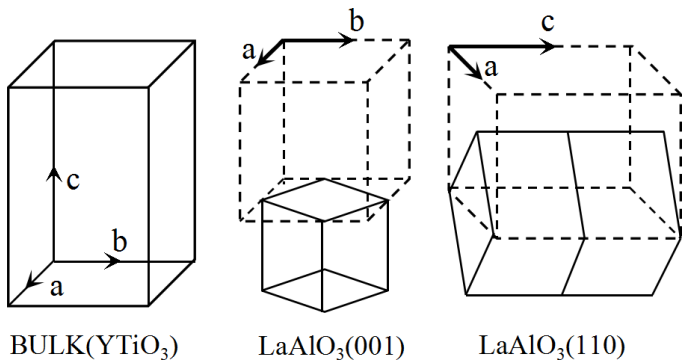}
\caption{A schematic diagram showing the epitaxial growth relationship between the YTiO$_3$ films and LaAlO$_3$ substrates. Left: Bulk YTiO$_3$ and its crystal axes. Middle: YTiO$_3$ films grown on the LaAlO$_3$ ($001$) substrates. Right: YTiO$_3$ films grown on the LaAlO$_3$ ($110$) substrates. Since the lattice constant of YTiO$_3$ along the $a$-axis instead of the $b$-axis is more close to the LaAlO$_3$ ($110$), the grown orientation of YTiO$_3$ on the LaAlO$_3$ ($110$) will be along the $b$-axis. The solid and dashed lines correspond to the LaAlO$_3$ substrates and YTiO$_3$ films, respectively.}
\label{sketch}
\end{figure}

The crystal structure of YTiO$_3$ bulk is orthorhombic, with the space group \textit{Pbnm}. The lattice constants are: $a=5.358$ \AA{}, $b=5.696$ \AA{}, and $c=7.637$ \AA{}. Such a minimum unit cell consists of four formula units. To simulate the effect of the epitaxial strain induced by the LaAlO$_3$ ($110$) substrate, the particular lattice constants along the $a$- and $c$-axes are fixed to match the ($100$) and ($1-10$) directions of LaAlO$_3$, namely, $a=5.366$ ($3.794\times\sqrt{2})$ \AA{} and $c=7.588$ ($3.794\times{2}$) \AA{}. Note that this epitaxial growth mode is different from the usual mode along the ($001$)-direction, in which case the in-plane lattice constants $a$ and $b$ are fixed to fit the [$001$] surface of substrate, as compared in Fig.~\ref{sketch}. Once the strain effect is considered, the out-of-plane lattice constant (e.g. the $b$-axis here) is optimized to search for the equilibrium one.

Our density-functional theory (DFT) calculations are performed using the projector-augmented wave (PAW) potentials \cite{Blochl:Prb2,Kresse:Prb99} as implemented in the Vienna \emph{ab} initio Simulation Package (VASP).\cite{Kresse:Prb,Kresse:Prb96} The electronic correlation is treated using the generalized gradient approximation (GGA) method with Hubbard $U$.\cite{Perdew:Prl} The Dudarev implementation\cite{Dudarev:Prb} is adopted with an on-site Coulomb interaction $U_{\rm eff}=U-J=3.2$ eV applied to the $3d$ electrons of Ti.\cite{Sawada:Prb1} The valence states include $4s4p5s4d$, $3d4s$ and $2s2p$ for Y, Ti and O, respectively. The atomic positions are fully optimized as the Hellman-Feynman forces are converged to less than $10$ meV/\AA{}. All calculations have been carried out using the plane-wave cutoff energy of $500$ eV and a $7\times7\times5$ Monkhorst-Pack $k$-point mesh centered at $\Gamma$ grid in combination with the tetrahedron method.\cite{Blochl:Prb}

%%------------------------------------------------------------------------
%

Before investigating the effect of the epitaxial strain, we checked the properties of the ground state of YTiO$_3$ bulk. With GGA+$U$, a Mott-insulator with ground FM state is obtained. The calculated band gap of $1.5$ eV is slightly overestimated compared to the experimental value $1.2$ eV\cite{Okimoto:Prb} and the local magnetic moment of Ti is $0.88$ $\mu_{\rm B}$, which agrees quite well with the experimental results.\cite{Garrett:Mrb}

\begin{figure}
\centering
\includegraphics[width=0.5\textwidth]{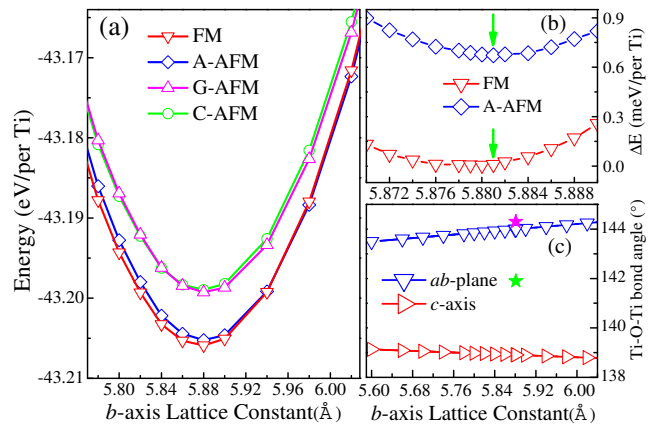}
\caption{(Color online) (a) Energies for different magnetic orders as a function of the $b$-axis lattice constant. (b) The energy difference between the A-AFM and FM as a function of the $b$-axis lattice constant. The lowest energy is taken as reference point. The equilibrium position is indicated by the green arrow. (c) The Ti-O-Ti bond angle in the \emph{ab}-plane and along the $c$-axis respectively for the ground FM state. For comparison, the \emph{ab}-plane and $c$-axis bond angles for bulk YTiO$_3$ are shown as the magenta and green stars, respectively.}
\label{energy}
\end{figure}

Next, the DFT calculations with the epitaxial strain were performed. As stated before, by fixing the lattice constants of $a$-axis and $c$-axis, the lattice constant along the $b$-axis is tuned from $5.6$ \AA{} to $6.1$ \AA{} to search for the equilibrium one. In our calculations, the internal atomic positions are relaxed with magnetism under each given lattice framework to obtain the optimal lattice structure for calculating accurate energies. To determine the ground state, several possible magnetic orders have been tested, which are FM, A-AFM, G-AFM, and C-AFM. As shown in Fig.~\ref{energy}(a), it is clearly seen that the FM and A-AFM have much lower energies than the C-AFM and G-AFM, while the energies of the FM and A-AFM states are very close. By zooming in on the energy curves of FM and A-AFM around the lowest position, as shown in Fig.~\ref{energy}(b), it is found that the minimum energy appears at $b$=$5.881$ \AA{} with the FM order. The optimized lattice constant along the $b$-axis for the A-AFM is 
also $5.881$ \AA{}, implying that the magnetostriction along the $b$-axis is negligible in YTiO$_3$. The energy difference between the FM and A-AFM at $5.881$ \AA{} is only about $0.66$ meV/per Ti. Although this energy difference is very small, the FM order is robust in a large range of $b$-axis lattice constant from $5.6$ \AA{} to $5.94$ \AA{}. Moreover, the ground state does not change as the Hubbard parameter $U_{\rm eff}$ changes from $0$ eV (pure GGA) to $5$ eV, implying that FM state is stable and not parameter-sensitive, although the change of the parameter $U$ can have a significant effect on magnetic energies.\cite{Fennie:Prl} The calculated FM ground state is the same as in the bulk YTiO$_3$, which agrees well with the experimental observation,\cite{Chae:Apl} but is totally different from the ($001$)-oriented case.

According to our previous study,\cite{Huang:Jap} the A-AFM is predicted to emerge against the original FM state when the YTiO$_3$ films are epitaxially grown on the ($001$)-oriented LaAlO$_3$ substrates, although the A-AFM does not exist in any bulks of titanic oxides. By interpreting the change of octahedral Ti-O-Ti bond angles, the origin of such an A-AFM order can be understood as the decreased bond angle ($\sim4.7$$^{\circ}$) in the \emph{ab}-plane and increased bond angle ($\sim1.8^{\circ}$) along the $c$-axis comparing with the original bulk values, since a more bending (straight) Ti-O-Ti bond prefers the FM (AFM) correlation according to the bulk's phase diagram.\cite{Okimoto:Prb,Katsufuji:Prb,Komarek:Prb,Mochizuki:Njp} Based on this scenario, we can also analyse the Ti-O-Ti bond angles in these ($110$)-oriented YTiO$_3$ films. As shown in Fig.~\ref{energy}(c), in contrast to the ($001$)-oriented case, the bond angle decreases for $3^{\circ}$ along the $c$-axis comparing with the bulk one. Therefore, 
the decreased bond angle can stabilize the FM correlation along the $c$-axis, against the possible A-AFM order. In the \emph{ab}-plane, the bond angle also deceases a little bit ($\sim0.4^{\circ}$), which will not alter the original FM order. On top of it, this variation of the Ti-O-Ti bond angle also agrees with the Goodenough-Kanamori rules qualitatively, which states that superexchange interactions are AFM where the two cations' orbitals (Ti here) overlap the same p orbital of a shared anion (O here) as in a $180^{\circ}$ cation-anion-cation bridge, but they are FM as in a $90^{\circ}$ cation-anion-cation bridge.\cite{Goodenough:Pr,Kanamori:Jpcs}

\begin{figure}
\centering
\includegraphics[width=0.45\textwidth]{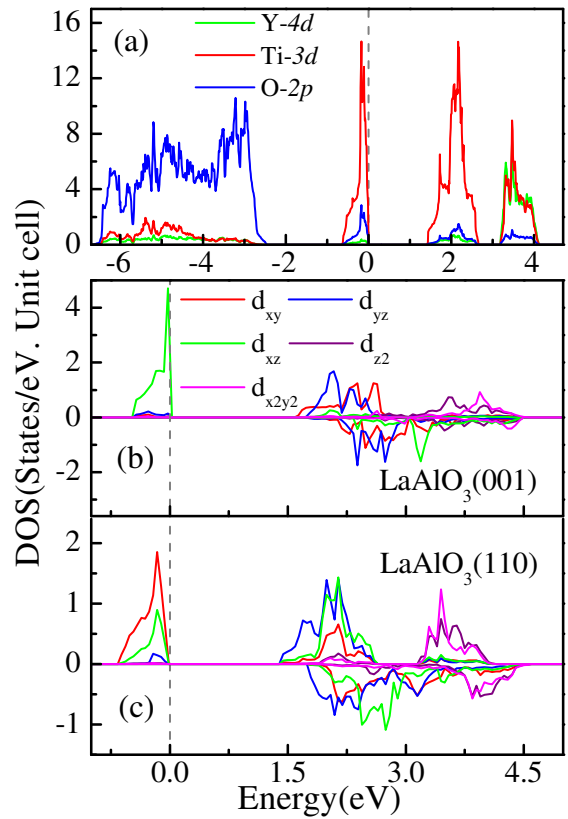}
\caption{(Color online) (a) The atomic-resolved PDOS of YTiO$_3$ films grown on the ($110$)-oriented LaAlO$_3$ substrates. The orbital-resolved PDOS of Ti-$3d$ orbitals on (b) the ($001$)-oriented LaAlO$_3$ substrates and (c) the ($110$)-oriented LaAlO$_3$ substrates. The gray dotted line indicates the Fermi energy.}
\label{orbital}
\end{figure}

It is well known that the magnetism often couples strongly with the $3d$ orbital ordering in the perovskite oxides, especially in $R$TiO$_3$ compounds.\cite{Zhou:Jpcm,Goodenough:Jmc,Mochizuki:Njp} Hence, to further understand the origin of this orientation-dependent magnetism in YTiO$_3$ films, the projected density of states (PDOS) are analyzed, as shown in Fig.~\ref{orbital}. Fig.~\ref{orbital}(a) shows the PDOS for Y's $4d$, Ti's $3d$, and O's $2p$ orbitals, indicating that the bands near the Fermi level are mostly contributed by Ti's $3d$ orbitals, while Y and O do not contribute significantly to these bands, as expected. Then, these Ti's $3d$ orbitals near the Fermi level are further split by the crystal field from the local octahedral framework. The orbital-resolved PDOS of Ti's $3d$ orbitals for the ($001$)- and ($110$)-orientated films are shown in Fig.~\ref{orbital}(b-c) for comparison. It is clearly seen that the contribution of each $3d$ orbital is different for these two cases. For the ($001$) 
one, the bands near the Fermi level are mainly occupied by the $d_{xz}$ orbitals for the selected Ti, while for its in-plane nearest-neighbor Ti's, they are mainly contributed by the $d_{yz}$ orbitals (not shown), rending the $d_{xz}/d_{yz}$ type orbital ordering pattern. In contrast, for the ($110$) one, the largest contribution to the occupied bands is from the $d_{xy}$ orbital, and the $d_{xz}$ orbital is also partially involved for the selected Ti. For its in-plane nearest-neighbor Ti's, the minority orbital changes to $d_{yz}$ and the majority orbital remains $d_{xy}$ (not shown). Such a difference links closely to the variable distortions of octahedral TiO$_6$, e.g. the Jahn-Teller modes, which have a significant effect on electronic structure as well as magnetism.\cite{Zhou:Prb}

\begin{figure}
\centering
\includegraphics[width=0.5\textwidth]{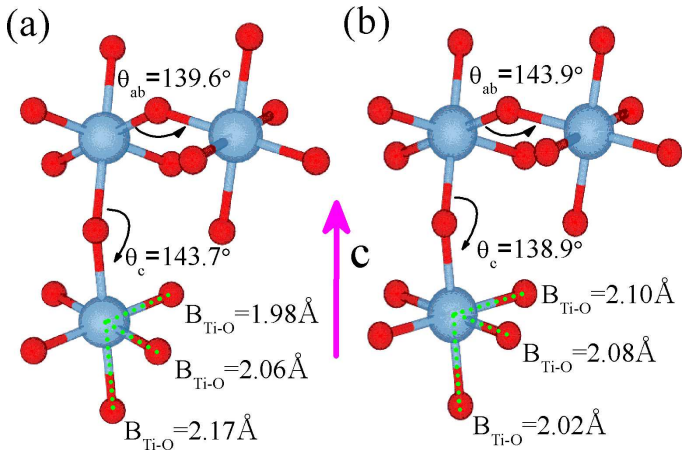}
\caption{(Color online) The optimized Ti-O-Ti bond angles ($\theta$) and Ti-O bonds lengths ($B_{\rm Ti-O}$) in strained YTiO$_3$ films for (a) the ($001$)-oriented one and (b) the ($110$)-oriented one.}
\label{bond}
\end{figure}

To confirm this argument, the Ti-O-Ti bond angles and Ti-O bond lengths in YTiO$_3$ films are calculated, as shown in Fig.~\ref{bond}, which exhibit some remarkable contrasts between these two orientations. First, the bond lengths for three Ti-O bonds in TiO$_6$ octahedra are distinct from each other. They are $1.98$ \AA{}, $2.06$ \AA{} (in \emph{ab}-plane) and $2.17$ \AA{} (along $c$-axis) for the ($001$)-oriented one, but $2.10$ \AA{}, $2.08$ \AA{} (in \emph{ab}-plane) and $2.02$ \AA{} (along $c$-axis) for the ($110$)-oriented one. For the ($001$) case, the longest Ti-O bond along the $c$-axis, associating with the Jahn-Teller $Q_3$ mode, will split the $t_{\rm 2g}$ triplet into higher $d_{xy}$ and low-lying $d_{xz}$/$d_{yz}$. The disproportion of in-plane Ti-O bond lengths, namely the Jahn-Teller $Q_2$ mode, will further split the $d_{xz}$/$d_{yz}$, giving rise to the orbital ordering as shown in Fig.~\ref{orbital}(b). For the ($110$) case, the out-of-plane bond length is the shortest, giving rise to a 
negative value for the Jahn-Teller $Q_3$ mode. As a result, the $d_{xy}$ is mostly preferred. However, the disproportion of Ti-O bond lengths are not so strong in this ($110$) case, thus partial contribution from the $d_{xz}$ (or $d_{yz}$) orbital can be mixed in considering the Jahn-Teller $Q_2$ mode. Furthermore, the Ti-O-Ti bonds are straighter in the \emph{ab}-plane for the ($110$)-oriented one compared with the ($001$)-oriented case, which is beneficial for electron hopping in-plane and thus gives rise to a wider bandwidth for occupied bands (Fig.~\ref{orbital}(b-c)).

In addition, Fig.~\ref{orbital} also shows that the YTiO$_3$ films retain insulating behavior regardless of the strain orientations. The band gap for the ($110$)-orientation is about $1.4$ eV, which is a little smaller than the ($001$) case.

\begin{table}
\caption{The lattice mismatch between the YTiO$_3$ films and LaAlO$_3$ substrates. Here ``+" means tensile strain and ``-" denotes compressive strain. The lattice constants are in unit of \AA{}.}

\begin{tabular*}{0.48\textwidth}{@{\extracolsep{\fill}}llllr}
\hline \hline
 & Bulk & ($001$) & ($110$)\\
\hline
$a$ & $5.358$ & $5.366$ ($+0.15\%$) & $5.366$ ($+0.15\%$)\\
$b$ & $5.696$ & $5.366$ ($-5.79\%$) & $5.881$\\
$c$ & $7.637$ & $8.250$ & $7.588$ ($-0.64\%$)\\
\hline \hline
\end{tabular*}
\label{mismatch}
\end{table}

At last, lattice mismatch should also be considered, which plays an important role to govern physical properties of thin films.\cite{Lee:Prl10} The lattice mismatches between the YTiO$_3$ films and LaAlO$_3$ substrates are summarized in Table.~\ref{mismatch}. It is found that the lattice mismatch for the YTiO$_3$ films grown on the ($001$)-oriented LaAlO$_3$ substrates is rather high and anisotropic, which reaches $-5.79$\% along the $b$-axis and +$0.15$\% along the $a$-axis. While using the ($110$)-oriented LaAlO$_3$ substrates, the lattice mismatch is greatly reduced, which is only about -$0.64$\% and +$0.15$\% along the $c$- and $a$- axes, respectively. Reducing the lattice mismatch is beneficial for reducing the strain of highly distorted orthorhombic YTiO$_3$ films on nearly cubic LaAlO$_3$ substrates, which restores the original FM order as the ground state.

%%------------------------------------------------------------------------
%
In summary, we have studied the effects of epitaxial strain on the magnetic states in the YTiO$_3$ films growning on the LaAlO$_3$ substrates with different orientations, e.g. ($110$) and ($001$). Our results show that the ground magnetic states for these two cases are totally different. The ferromagnetic order is the most stable state for the ($110$)-oriented film, in contrast to the A-type antiferromagnetic order on the ($001$) substrate. This orientation-dependent magnetism links closely to the variation of Ti-O-Ti bond angles, as well as the bond lengths. As such orientation-dependent magnetism is accompanied by the particular $3d$ orbital ordering, the distortions of TiO$_6$ octahedra play an important role in this system.

\begin{acknowledgments}
Work was supported by the 973 Projects of China (Grant No. 2011CB922101), Natural Science Foundation of China (Grant Nos. 11274060 and 51322206 and 51172044 and 51471085).

\end{acknowledgments}

\bibliographystyle{apsrev4-1}
\bibliography{ref}
\end{document}